\documentclass[conference]{IEEEtran}

\usepackage{cite}
\usepackage[pdftex]{graphicx}
\usepackage[cmex10]{amsmath}
\usepackage[caption=false,font=footnotesize]{subfig}

\usepackage{rmsbrt_math}
\usepackage{rmsbrt_IBC}
\usepackage{rmsbrt_algorithm}
\usepackage{rmsbrt_sets}

\usepackage[T1]{fontenc}
\usepackage[utf8]{inputenc}

\begin{document}

\title{Interference Alignment-Aided Base Station Clustering using Coalition Formation}

\author{\IEEEauthorblockN{Rasmus Brandt, Rami Mochaourab, and Mats Bengtsson}
\IEEEauthorblockA{Dept. of Signal Processing, ACCESS Linn\ae{}us Centre, KTH Royal Institute of Technology, Stockholm, Sweden}}

\maketitle

%!TEX root = asilomar.tex

\begin{abstract}
Base station clustering is necessary in large interference networks, where the channel state information (CSI) acquisition overhead otherwise would be overwhelming. In this paper, we propose a novel long-term throughput model for the clustered users which addresses the balance between interference mitigation capability and CSI acquisition overhead. The model only depends on statistical CSI, thus enabling long-term clustering. Based on notions from coalitional game theory, we propose a low-complexity distributed clustering method. The algorithm converges in a couple of iterations, and only requires limited communication between base stations. Numerical simulations show the viability of the proposed approach.
\end{abstract}

%!TEX root = asilomar.tex

\section{Introduction}
Multicell coordinated precoding \cite{Gesbert2010}, of which interference alignment (IA) \cite{Cadambe2008} is a particular kind, is a promising technique for achieving high cell-edge rates in future 5G networks. Large sum rate gains are possible by exploiting the spatial separability of the receivers, but this requires channel state information at the transmitters (CSI-T). For frequency-division duplex systems, the feedback load for obtaining the CSI-T scales quadratically with the number of transmitters partaking in the coordinated operation. This holds both for analog and digital feedback. At finite signal-to-noise ratios (SNRs), only strong interferers matter however. Due to the CSI acquisition overhead incurred, it is therefore not worthwhile to coordinate all transmitters in large networks \cite{Lozano2013}. Base station clustering is an approach for balancing the substantial interference mitigation ability of large clusters with the correspondingly high CSI acquisition overhead. Existing work on base station clustering using IA includes \cite{Chen2014}, where a graph partitioning problem was formulated based on fixed edge weights calculated from statistical CSI. Contrary to our work, it is however unclear if the model can modified to directly incorporate the effect of CSI acquisition overhead. In \cite{Peters2012}, the CSI acquisition overhead was modelled using a rate pre-log factor, which was then used to formulate heuristic algorithms for base station grouping in orthogonal time slots. We use a similar overhead model, but assume full spectral reuse. Static base station clustering for hexagonal cells was studied in \cite{Tresch2009}. In \cite{Pantisano2013}, a coalitional game in partition form was used for femtocell clustering. The CSI acquisition overhead was modelled in terms of transmission power which, due to a total power constraint, limited the amount of power left for data transmission.

Our contribution is a novel model for the users' long-term throughputs assuming IA precoding within the clusters, and an accompanying base station clustering algorithm based on coalition formation \cite{Thrall1963}. The proposed throughput model takes into account the trade-off between interference mitigation ability and CSI acquisition overhead. By formulating a coalitional game in partition form \cite{Thrall1963}, the low-complexity distributed coalition formation algorithm is devised. Although the system is developed assuming IA, the resulting base station clustering can be combined with any other precoding method. Despite its low complexity, the proposed approach performs remarkably well in our numerical simulation.

%!TEX root = asilomar.tex

\begin{figure*}[!t]
    \centering
    \includegraphics[width=\textwidth]{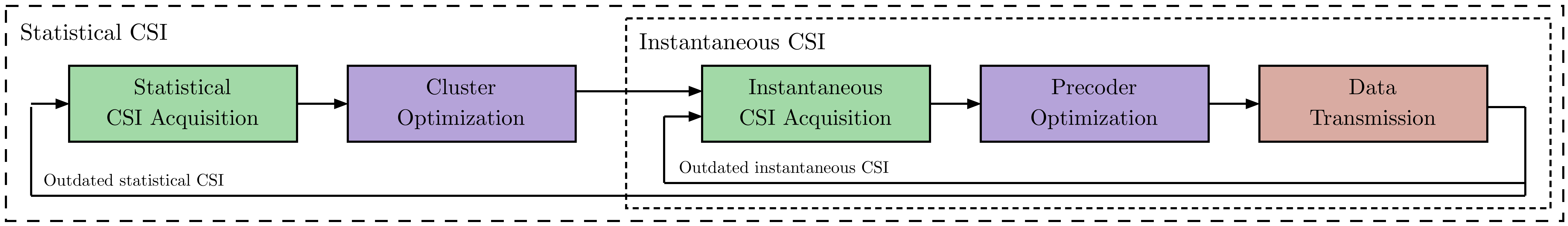}
    \caption{Block diagram of proposed system. Note that the statistical CSI acquisition does not necessarily need to be a discrete block, but could be performed concurrently with system operation based on instantaneous CSI. As a rough approximation for intermediate MS mobility, the statistical CSI must be re-estimated on the order of seconds, whereas the instantaneous CSI must be re-estimated on the order of milliseconds.} \label{fig:block_diagram}
\end{figure*}

\section{System Model}
We study a frequency-division duplex network composed of $K$ base stations (BSs), each serving a single mobile station (MS). We index the BS/MS-pairs (which we will call \emph{users}) using the set $\setK = \{ 1, 2, \ldots, K \}$. All BSs use the same time and frequency resources for serving their respective MSs, and the interference channel (IC) is thus used as the model for the physical layer. The interference created is handled by coordinated precoding, enabled by the sharing of CSI-T between cooperating BSs. The means of BS cooperation is encoded in a \emph{coalition structure} \cite{Thrall1963}:
\begin{definition}[Coalition Structure] \label{def:coalition_structure}
    A coalition structure $\setS = \{ \setC_1, \setC_2, \ldots, \setC_P \}$ is a partition of $\setK$ into disjoint sets called coalitions, such that $\varnothing \neq \setC_p \subseteq \setK$ and $\setC_p \cap \setC_q = \varnothing$ for all $\setC_p, \setC_q \in \setS, p \neq q$. Further, $\bigcup_{p = 1}^P \setC_p = \setK$. For a user $k \in \setC_p$, we let $\Pi_k(\setS) = \setC_p$ map to its coalition.
\end{definition}
The channel between BS $l$ and MS $k$ is denoted as $\Hkl~\in~\complexnumbers^{N_k \times M_l}$. We consider the case when the elements of $\Hkl$ are i.i.d. $\mathcal{CN} \left( 0, \gamma_{kl} \right)$, where $\gamma_{kl}$ describes the large-scale fading. BS $k$ sends a signal $\xk \sim \mathcal{CN} \left( 0, \matI_{d_k} \right)$ to MS $k$, where $d_k$ is the number of spatial data streams. For spatial interference suppression, BS $k$ uses a linear precoder $\Vk \in \complexnumbers^{M_k \times d_k}$ and MS $k$ uses a linear receive filter $\Uk \in \complexnumbers^{N_k \times d_k}$. The received signals are thus modelled as
\begin{equation*}
    \begin{split}
        \yk = &\underbrace{\vphantom{\sum_{l \in \Pi_k(\setS) \setminus \{ k \}}}\Uk^\herm \Hkk \Vk \xk}_\text{desired signal} + \underbrace{\Uk^\herm \hspace{-5mm} \sum_{l \in \Pi_k(\setS) \setminus \{ k \}} \hspace{-5mm} \Hkl \Vl \xl}_\text{intracoalition interference} \\
        + &\underbrace{\Uk^\herm \hspace{-3mm} \sum_{l \in \setK \setminus \Pi_k(\setS)} \hspace{-3mm} \Hkl \Vl \xl}_\text{intercoalition interference} + \underbrace{\vphantom{\sum_{l \in \setK \setminus \Pi_k(\setS)}}\Uk^\herm \zk}_\text{filtered noise}, \; \forall \, k \in \setK.
    \end{split}
\end{equation*}
It is our goal to design the coalition structure such that a good trade-off between interference mitigation and CSI acquisition overhead is achieved. In order to do so, we first derive a model for the long-term throughputs of the MSs.

\subsection{Long-Term Spectral Efficiency}
The users in a coalition cooperate to completely remove all intracoalition interference.
\begin{assumption}[Intracoalition CSI Sharing] \label{ass:cooperation}
    The users in a coalition share their local \emph{instantaneous CSI} such that user $k$ has access to $\{ \Hkl \}_{l \in \Pi_k(\setS)}$ and $\{ \Hlk \}_{l \in \Pi_k(\setS)}$. All users have access to full local \emph{statistical CSI} (i.e. $\{ \gamma_{kl} \}_{l \in \setK}$ and $\{ \gamma_{lk} \}_{l \in \setK}$).
\end{assumption}
\begin{assumption}[Intracoalition Interference Alignment] \label{ass:IIA}
    The precoders $\{ \Vk \}_{k \in \setK}$ and receive filters $\{ \Uk \}_{k \in \setK}$ satisfy:
    \begin{align}
        \Uk^\herm \Hkl \Vl &= \matO, \quad \forall \, k, l \in \Pi_k(\setS), l \neq k, \label{eq:IA_condition_1} \\
        \textnormal{rank} \left( \Uk^\herm \Hkk \Vk \right) &= d_k, \quad \forall \, k \in \Pi_k(\setS). \label{eq:IA_condition_2}
    \end{align}
\end{assumption}
Note that \eqref{eq:IA_condition_1}--\eqref{eq:IA_condition_2} can only be satisfied if IA is feasible \cite{Yetis2010} for all coalitions in $\setS$. IA is well-known to achieve the maximum degrees of freedom\footnote{The degrees of freedom of the interference channel can be seen as the number of interference-free data streams that can successfully be transmitted in the network, or equivalently as the pre-log factor of the sum capacity.} of the interference channel \cite{Cadambe2008} when there is no uncoordinated interference. Unless it is feasible for all users to join a single coalition, the so called \emph{grand coalition} $\setS^\text{grand} = \{ \setK \}$, the remaining non-aligned intercoalition interference will reduce the degrees of freedom to zero. With a well-designed coalition structure however, a major portion of the interference power will correspond to intracoalition interference, which can be aligned and removed using IA. The resulting high-SNR performance will then only be limited by intercoalition interference. To formulate the corresponding long-term spectral efficiencies, we require some conditions on the IA solution.
\begin{assumption}[Properties of IA Solution] \label{ass:IA_properties}
    For the IA solution satisfying Assumption~\ref{ass:IIA}, the following holds for all $k \in \setK$:
    \begin{enumerate}
        \item[\textbf{A3.1}] $\Uk$ and $\Vk$ do not depend on $\Hkl$ for all $l \in \setK \setminus \Pi_k(\setS)$
        \item[\textbf{A3.2}] $\Vk$ does not depend on the direct channel $\Hkk$
        \item[\textbf{A3.3}] $\Vk$ is scaled semi-unitary (i.e. $\Vk^\herm \Vk = \frac{P_k}{d_k} \, \matI_{d_k}$)
        \item[\textbf{A3.4}] $\Uk$ is semi-unitary (i.e. $\Uk^\herm \Uk = \matI_{d_k}$) and designed such that $\ukn^\herm \Hkk \mathbf{v}_{k,m} = 0$ for $m \neq n$
    \end{enumerate}
\end{assumption}
Assumption 3.1 does not contradict Assumption~\ref{ass:IIA}. Assumption 3.2 holds for any IA method that aims to only directly satisfy \eqref{eq:IA_condition_1}, while relying on \eqref{eq:IA_condition_2} being satisfied with probability 1. This is the case for e.g. the leakage minimization algorithm \cite{Gomadam2011}, when applied to i.i.d. Rayleigh channels. Assumption 3.3 does not restrict the feasibility of IA \cite{Yetis2010}, and any IA solution can be transformed such that Assumption 3.4 holds \cite{ElAyach2012}.

Given an IA solution satisfying Assumption~\ref{ass:IA_properties}, the effective channels can be characterized using the following lemma (partially inspired by Lemma~1 in \cite{ElAyach2012}).
\begin{lemma} \label{lemma:effective_channels}
    The effective desired channel for the $n$th stream of MS $k$ is
    \begin{equation} \label{eq:barhkn_distribution}
        \bar{h}_{k,n} = \ukn^\herm \Hkk \vkn \sim \mathcal{CN} \left( 0, \gamma_{kk} P_k/d_k \right).
    \end{equation}
    The intercoalition effective interfering channel between the $m$th stream of BS $l$ to the $n$th stream of MS $k$ is
    \begin{equation} \label{eq:tildehklnm_distribution}
        \tilde{h}_{kl,nm} = \ukn^\herm \Hkl \vlm  \sim \mathcal{CN} \left( 0, \gamma_{kl} P_l/d_l \right).
    \end{equation}
\end{lemma}
\begin{IEEEproof}
    Due to the bi-unitary invariance of $\Hkk$ \cite{RandomMatrixTheory}, together with Assumptions 3.2 and 3.3, each element of $\Hkk \vkn$ is i.i.d. $\mathcal{CN} \left( 0, \gamma_{kk} P_k/d_k \right)$. Due to Assumption 3.4, $\ukn$ is orthonormal and independent of $\Hkk \vkn$, and together with the bi-unitary invariance of $\Hkk \vkn$, \eqref{eq:barhkn_distribution} follows. Equation \eqref{eq:tildehklnm_distribution} follows similarly, except that Assumption 3.1 is applied for the independence between $\ukn$, $\Hkl$ and $\vlm$.
\end{IEEEproof}
Following Assumption~\ref{ass:IIA}, the intracoalition interference is cancelled and the resulting received signal for the $n$th stream of MS $k$ can be written as
\begin{equation} \label{eq:received_signal_after_assumptions}
    \ykn = \bar{h}_{k,n} \xkn + \sum_{l \in \setK \setminus \Pi_k(\setS)} \sum_{m=1}^{d_l} \tilde{h}_{kl,nm} \xjlm + \bar{z}_{k,n},
\end{equation}
where $\bar{z}_{k,n} = \ukn^\herm \zk \sim \mathcal{CN} \left( 0, \sigma_k^2 \right)$. The long-term \emph{spectral efficiency} for MS $k$ is now given by the following theorem.

\begin{figure*}[t!]
    \begin{equation}
        r_{k,n}(\setS) = \mathbb{E}_{\bar{h}_{k,n}} \log \left( 1 + \frac{\abssq{\bar{h}_{k,n}}}{\mathbb{E}_{\{ \tilde{h}_{kl,nm} \}} \sum_{l \in \setK \setminus \Pi_k(\setS)} \sum_{m=1}^{d_l} \abssq{\tilde{h}_{kl,nm}} + \sigma_k^2} \right) = \mathbb{E}_{\bar{h}_{k,n}} \log \left( 1 + \frac{\abssq{\bar{h}_{k,n}}}{\sum_{l \in \setK \setminus \Pi_k(\setS)} \gamma_{kl} P_l + \sigma_k^2} \right) \label{eq:long-term_stream_rates_def}
    \end{equation}
    \hrule
\end{figure*}

\begin{theorem} \label{theorem:long-term_spectral_efficiency}
    If the intercoalition interference is treated as noise in a nearest neighbour decoder, the long-term spectral efficiency for MS $k$ is
    \begin{equation*}
        r_k(\setS) = d_k e^{1/\rho_k(\setS)} E_1 \left( {1/\rho_k(\setS)} \right),
    \end{equation*}
    where $\rho_k(\setS) = \frac{\gamma_{kk} P_k/d_k}{\sum_{l \in \setK \setminus \Pi_k(\setS)} \gamma_{kl} P_l + \sigma_k^2}$ is the per-stream SNR and $E_1 \left( \xi \right) = \int_{\xi}^\infty t^{-1} e^{-t} \, \mathrm{d}t$ is the exponential integral.\footnote{The function $E_1(\xi)$ can be calculated numerically, e.g. by summing its truncated power series expansion \cite[5.1.11]{AbramowitzStegun}.}
\end{theorem}
\begin{IEEEproof}
    Under Assumption~\ref{ass:cooperation}, $\{ \bar{h}_{k,n} \}_{n = 1, \ldots, d_k}$ are known to MS $k$, but $\{ \tilde{h}_{kl,nm} \}_{l \in \setK \setminus \Pi_k(\setS), n = 1, \ldots, d_k, m = 1, \ldots, d_l}$ are unknown. Due to the construction of the decoder, and since $\xk$ is Gaussian, the intercoalition interference plays the role of additional Gaussian noise \cite{Lapidoth1996}. Given Lemma~\ref{lemma:effective_channels}, the achievable spectral efficiency for the $n$th stream of MS $k$ can then be written as in \eqref{eq:long-term_stream_rates_def}, at the top of the page. Integrating by parts, it can be shown that the right hand side of \eqref{eq:long-term_stream_rates_def} equivalently is $r_{k,n}(\setS) = e^{1/\rho_k(\setS)} E_1 \left( 1/\rho_k(\setS) \right)$, and the result follows.
\end{IEEEproof}

\subsection{CSI Acquisition Overhead and Long-Term Throughput}
Under block fading, with coherence time $T_c \, [\text{s}]$ and coherence bandwidth $W_c \, [\text{Hz}]$, the coherence block length is $L^\text{coh} = T_cW_c$ \cite{Lozano2013}. Within each coherence block, the instantaneous CSI must be re-acquired (cf. Fig.~\ref{fig:block_diagram}). As a model for the corresponding cost\footnote{Relative to the cost of instantaneous CSI acquisition, we assume that the cost of statistical CSI acquisition is negligible.} in terms of lost temporal degrees of freedom, we multiply the spectral efficiency with a pre-log factor $\alpha(\setS)$, such that the long-term \emph{throughput} for MS $k$ is
\begin{equation} \label{eq:throughput}
    t_k \left( \setS \right) =
    \begin{cases}
        \alpha \left( \setS \right) r_k \left( \setS \right) & \text{if IA is feasible for $\Pi_k(\setS)$} \\
        0 & \text{otherwise}
    \end{cases}.
\end{equation}
With $L_{\setC_p}^\text{CSI}$ being the number of symbol intervals required for CSI acquisition in coalition $\setC_p$, we model the pre-log factor as $\alpha(\setS) = 1 - \frac{\sum_{p=1}^P L_{\setC_p}^\text{CSI}}{L^\text{coh}}$ since, in order to not interfere, the coalitions must perform CSI acquisition orthogonally. Assuming pilot-assisted channel training and analog feedback \cite{ElAyach2012}, the \emph{minimum} number of symbol intervals needed for the CSI-T acquisition within coalition $\setC_p$ can be modelled\footnote{This is under the assumption that $\sum_{k \in \setC_p} M_k \geq N_l, \; \forall \, l \in \setC_p$, such that each channel matrix fed back can be successfully jointly estimated at the BSs in $\setC_p$ \cite{ElAyach2012}. A sufficient condition for this to hold is if $M_k \geq N_l, \; \forall \, k, l \in \setK$.} as \cite{Mochaourab2015submitted}:
\begin{equation*}
    L_{\setC_p}^\text{CSI} = \underbrace{\sum_{k \in \setC_p} M_k}_\text{DL training} + \underbrace{\sum_{k \in \setC_p} N_k}_\text{UL training} + \underbrace{\sum_{k \in \setC_p} \sum_{l \in \setC_p} M_l}_\text{analog feedback} + \underbrace{\sum_{k \in \setC_p} d_k.}_\text{DL effective ch. training}
\end{equation*}
For large coalitions, the dominating term in $L_{\setC_p}^\text{CSI}$ will be the analog feedback term, which grows quadratically in the cardinality of the coalition $\lvert \setC_p \lvert$.

Given IA feasibility, the grand coalition $\setS^\text{grand}$ will maximize $r_k \left( \setS \right)$ for all $k \in \setK$. Similarly, the coalition structure $\setS^\text{singletons} = \{ \{ 1 \}, \{ 2 \}, \ldots, \{ K \} \}$ of singleton coalitions will maximize $\alpha \left( \setS \right)$. Neither of these extremes is generally optimal, and our goal is thus to find a distributed and low-complexity trade-off between these. For this, we use the framework of coalitional games.

%!TEX root = asilomar.tex

\section{Coalitional Game}\label{sec:coalition_formation}
Coalitional games model cooperation between rational players \cite{Thrall1963}. A coalition between a set of players forms whenever they can take joint actions which lead to mutual benefits. Since the long-term throughput of each user depends on the whole coalition structure, it is suitable to study the coalitional game in partition form \cite{Thrall1963}. We formulate the game associated with our setting as $\langle \setK, \{b_k\}_{k\in \setK}, \{\tilde{t}_k\}_{k\in \setK}\rangle$. Each player in $\setK$ has a history set $\setH_k$ which includes the coalitions it has belonged to in the past, as well as a communication budget $b_k\in\mathbb{N}$ which limits the number of times it can communicate with other coalitions. With these definitions, we let the utility of a player $k$ in the game be
\begin{equation} \label{eq:restricted_utility}
\hspace{-0.1em} \tilde{t}_k(\setS;\setH_k,\eta_k) = \begin{cases}
   t_k(\setS)\hspace{-0.1em}&\text{if } \Pi_k(\setS) \notin \setH_k \text{ and } \eta_k \leq b_k  \\
   0       &\text{otherwise}\\
  \end{cases},\hspace{-0.1em}
\end{equation}
which is the long-term throughput $t_k(\setS)$ in \eqref{eq:throughput} with two restrictions: (i) a player does not profit by joining a coalition it has been member of before, and (ii) a player $k$ should not exhaust its communication budget $\eta_k \leq b_k$ where $\eta_k \in\mathbb{N}$ is the number of times player $k$ has already communicated with another coalition to ask for cooperation. 

\subsection{Coalition Formation}
Coalition formation describes the dynamics which lead to stable coalition structures. We will specify the elements which are required to formulate the coalition formation game by using a coalition formation model from \cite{Bogomolnaia2002}. In this model, only a single player is allowed to deviate and leave its coalition to join another. A player can join a new coalition only if all members of the coalition accept him. Such a coalition formation model has been used, e.g. in \cite{Saad2012} in the context of cognitive radio settings.

\begin{definition}[Deviation]\label{def:deviation}
A player $k \in \setK$ leaves $\Pi_k(\setS)$ to join $\setT \in \setS \cup \{ \varnothing \}$. Consequently, the coalition structure changes to $\setS_{\setT}$. We capture this change by the notation $\setS_{\setT}\overset{k}{\longleftarrow} \setS$.
\end{definition}

A deviation is admissible if the player as well as the members of the coalition it wants to join improve their payoff.
\begin{definition}[Admissible Deviation]\label{def:admissable}
A deviation $\setS_{\setT}~\overset{k}{\longleftarrow}~\setS$ is admissible if $\tilde{t}_k(\setS_\setT;\setH_k,\eta_k) > \tilde{t}_k(\setS;\setH_k,\eta_k),$ and $\tilde{t}_l(\setS_{\setT};\setH_l,\eta_l) \geq \tilde{t}_l(\setS;\setH_l,\eta_l), \text{ for all players } l \in \setT.$
\end{definition}
The rational players will pursue admissible deviations.

\begin{definition}[Individual Stability]\label{def:Individual_stability}
A coalition structure $\setS$ is individually stable if there exists no player $k\in\setK$ and coalition $\setT$ such that $\setS_{\setT}\overset{k}{\longleftarrow} \setS$ is admissible.
\end{definition}

\begin{algorithm}[t]
    \caption{\label{alg:coalition0} Coalition Formation for Base Station Clustering}
    \begin{algorithmic}[1]
    \Statex \textbf{Initialize}: $i = 1$, $\setS^{(1)} = \setS^\text{grand}$, $\setH_k = \varnothing$, $\eta_k = 0$, $k \in \setK$
    \Repeat 
    \Loop ~over $k$ sequentially from $\setK$
        \Loop{~in a greedy order over all $\setT \in \Psi_k(\setS^{(i)})$}
        \State Increment search factor $\eta_k = \eta_k + 1$
            \If{deviation $\setS_{\setT}\overset{k}{\longleftarrow} \setS$ is admissible}
                \State Update $\setS^{(i+1)} = \setS_{\setT}^{(i)}$
                \State Update $\setH_k = \setH_k \cup \Pi_k(\setS^{(i)})$
                \State Go to line 12
            \EndIf
        \EndLoop
        \State $i = i + 1$
    \EndLoop
    \Until{no player deviated}
    \end{algorithmic}
\end{algorithm}

Relying on the deviation model and stability concept we formulate the coalition formation algorithm in Algorithm~\ref{alg:coalition0}. The initial coalition structure is the grand coalition, but the algorithm could be initialized in any other coalition structure. Based on the current coalition structure $\setS^{(i)}$, player $k$ can calculate its acceptable coalitions as
\begin{equation*}
\Psi_k(\setS) = \{\setT \in \setS\cup \{ \varnothing \} \mid \tilde{t}_k(\setS_{\setT};\setH_k,\eta_k) > \tilde{t}_k(\setS;\setH_k,\eta_k)\}.
\end{equation*}
The utilities in the acceptable coalitions are stored in decreasing order such that player $k$ can greedily select coalitions which provide the highest utility first. After incrementing the index $\eta_k$ on line 4, player $k$ tests if joining the coalition $\setT$ is strictly profitable for him. If this is the case, it asks the members of coalition $\setT$ for permission to join. Here, we assume that the players are able to communicate for this purpose. Each player in the coalition can calculate its utility locally and communicate its decision with the others in the coalition. If all members of coalition $\setT$ accept player $k$, then the deviation is admissible. The coalition structure changes (line 6) after player $k$ leaves its coalition to join the new one. The history set of player $k$ is updated on line 7 to include the previous coalition it was member of. The coalition formation stops when no deviations are admissible.

\begin{theorem}
Algorithm \ref{alg:coalition0} converges to an individually stable coalition structure. The number of iterations is upper bounded by the minimum of $\sum_{k \in \setK} b_k$ and the $K$th Bell number.
\end{theorem}
\begin{IEEEproof}
Convergence is guaranteed due to the restrictions imposed through the history set and communication budget in \eqref{eq:restricted_utility}. Each player $k$ can change coalitions a maximum of $b_k$ times. Thus the maximum number of iterations is $\sum_{k \in \setK} b_k$. However, if the budget constraints are large enough, the worst case number of iterations is limited by the number of partitions of $\setK$, which is the $K$th Bell number \cite[Proof of Thm. 1]{Saad2012}. The stability result follows from iterating over all deviation possibilities until no admissible deviations remain.
\end{IEEEproof}

%!TEX root = asilomar.tex

\section{Performance Evaluation}
We evaluate the performance of the proposed approach by means of numerical simulations. The source code is made available at \cite{asilomar2015_reproducible_research}. The simulation parameters are inspired by 3GPP Case 2 \cite{TR25814}. We consider a scenario where $K$ BSs and $K$ MSs are placed uniformly at random in a square, whose dimensions are selected such that the average cell size is identical to that of a deployment where the BSs have hexagonal cells with an apothem of $250$ m \cite{TR25814}. A greedy user association algorithm is applied, associating one MS with each BS. We consider symmetric networks where all BSs have $M$ antennas, all MSs have $N$ antennas and are served $d = 1$ data streams each. We use the notation $(N \times M, d)^K$ to refer to such a network \cite{Yetis2010}. We assume a bandwidth of $10 \, \text{MHz}$ and a carrier frequency of $f_c = 2 \, \text{GHz}$, such that the pathloss model $\text{PL}_\text{dB} = 15.3 + 3.76 \log_{10} \left( \texttt{distance} [m] \right)$ is relevant \cite{TR25814}. Log-normal i.i.d. shadow fading with a standard deviation of $8$ dB is used for all channels, and the small-scale fading is i.i.d. Rayleigh fading. The coherence bandwidth is assumed to be $W_c = 300 \, \text{kHz}$, and the noise power spectral density is $-174$ dBm/Hz together with a receiver noise figure of $9$ dB.

\begin{figure}[t]
    \centering
    \includegraphics[width=\columnwidth]{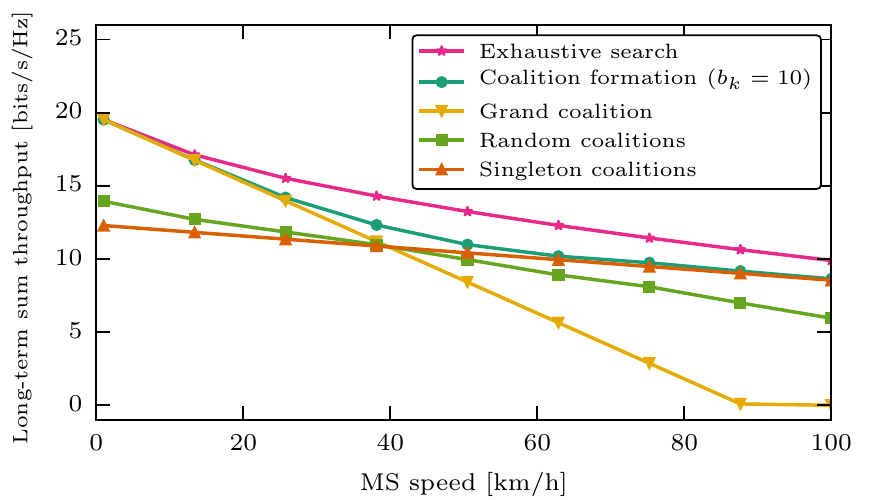} 
    \caption{Results with IA precoding in the $(6 \times 12, 1)^8$ IC.} \label{fig:small_network1-coherence-longterm_sumrate}
    \vspace{-1em}
\end{figure}

We generate $500$ independent realizations of the BS/MS positions and given each position realization, we generate $5$ independent small-scale fading realizations. System performance is evaluated in terms of the sum throughput $t_\text{sum} = \sum_{k \in \setK} t_k$. Algorithm~\ref{alg:coalition0} is compared to the k-means based algorithm in \cite{Chen2014}, since it outperforms the other algorithms of \cite{Chen2014}. For benchmarking, we also compare to the global optimum obtained by exhaustive search (for small $K$), the case of randomly generated coalitions, as well as the previously defined $\setS^\text{grand}$ and $\setS^\text{singletons}$ coalition structures. Three networks are studied:

\subsubsection{$(6 \times 12, 1)^8$ IC}
In this scenario, IA is feasible \cite{Yetis2010} for the grand coalition due to the excess of antennas. We vary the MS speed---thus varying $L^\text{coh}$---and provide the resulting long-term sum throughput in Fig.~\ref{fig:small_network1-coherence-longterm_sumrate}. At low speeds, the coherence block length is large, and the grand coalition is optimal since there this a lot of time to spend on CSI acquisition. At high speeds, the grand coalition throughput is zero, since there is not enough time to satisfy the CSI acquisition needs. Algorithm~\ref{alg:coalition0} is able to smoothly transition between the grand coalition and the singleton coalitions.

\begin{figure}[t]
    \centering
    \includegraphics[width=\columnwidth]{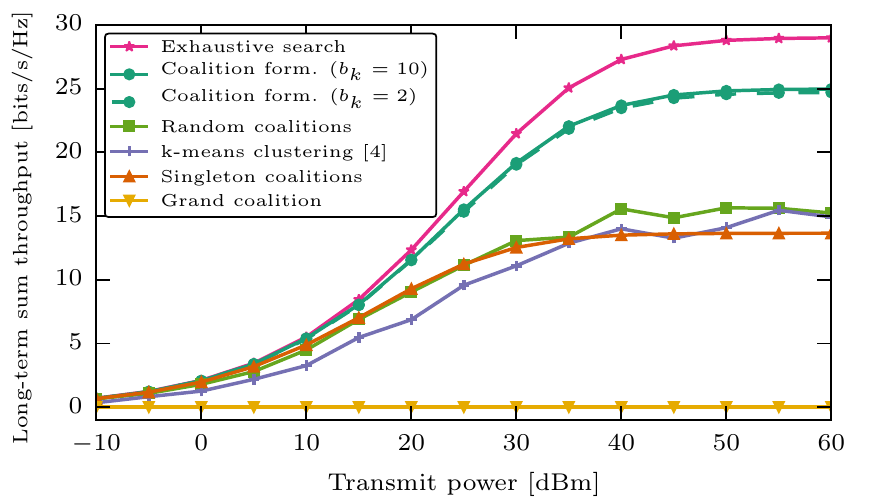} 
    \caption{Results with IA precoding in the $(2 \times 4, 1)^8$ IC.} \label{fig:small_network2-SNR-longterm_sumrate}
\end{figure}

\subsubsection{$(2 \times 4, 1)^8$ IC}
We contrast the previous scenario with an IA infeasible scenario, where the grand coalition will have zero sum throughput under IA precoding. For an MS speed of $30$ km/h \cite{TR25814}, corresponding to a coherence block length of $L^\text{coh} = 2700$ symbols, the long-term sum throughput is shown in Fig.~\ref{fig:small_network2-SNR-longterm_sumrate}. Algorithm~\ref{alg:coalition0} is outperforming all benchmarks except exhaustive search, whose shape it is able to follow. The k-means based algorithm of \cite{Chen2014} is not performing particularly well, the reason being that it may result in coalitions that are not strictly IA feasible.

\subsubsection{$(2 \times 4, 1)^{16}$ IC}
\begin{figure}[t]
    \centering
    \includegraphics[width=\columnwidth]{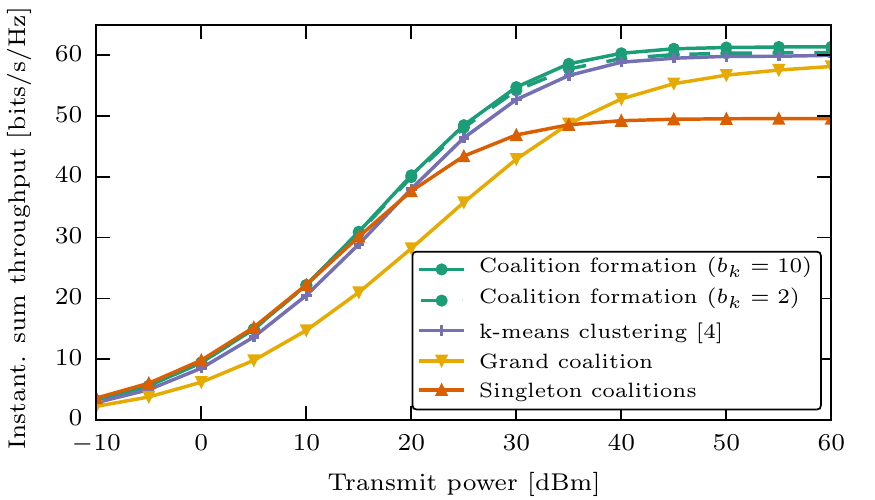} 
    \caption{Results with WMMSE precoding in the $(2 \times 4, 1)^{16}$ IC.} \label{fig:large_network-SNR-instantaneous_sumrate}
    \includegraphics[width=\columnwidth]{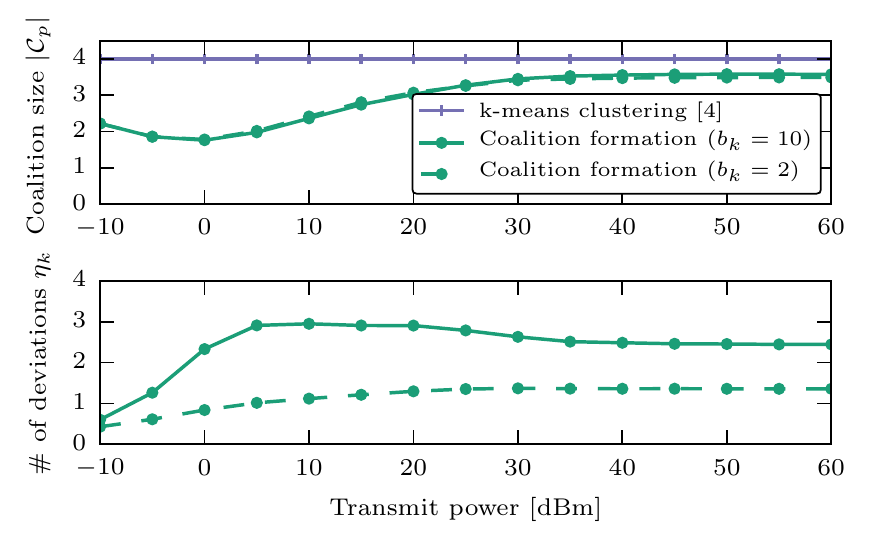}
    \caption{Average algorithm behaviour in the $(2 \times 4, 1)^{16}$ IC.} \label{fig:large_network-SNR-num_clusters-num_searches}
    \vspace{-1.2em}
\end{figure}

For this larger network, we evaluate performance in terms of instantaneous sum throughput after precoding. We use a modified version of the WMMSE algorithm \cite{Shi2011}, where the coloured noise of the intercoalition interference is replaced with white noise with equal average power. This can be shown to lead to a robust and convergent algorithm, but due to space reasons we leave out the details. The results are shown in Fig.~\ref{fig:large_network-SNR-instantaneous_sumrate}, where it can be seen that Algorithm~\ref{alg:coalition0} outperforms the benchmarks. Since the k-means based algorithm is not able to capture the CSI acquisition overhead being a function of the coalition sizes, it performs slightly worse. In Fig.~\ref{fig:large_network-SNR-num_clusters-num_searches}, we see that Algorithm~\ref{alg:coalition0} can adapt the coalition sizes, while still only using a couple of deviations.

%!TEX root = asilomar.tex

\section{Conclusions}
We have proposed a long-term throughput model for base station clustering using IA and an accompanying low-complexity distributed algorithm based on coalition formation. The numerical results show the importance of treating both spectral efficiency as well as CSI acquisition overhead when tackling the base station clustering problem.

\bibliographystyle{IEEEtran}
\bibliography{IEEEabrv,coordinated_precoding,rasmus_brandt,games}

\end{document}